\documentclass[nofootinbib,twocolumn,showpacs,preprintnumbers,pre,aps,superscriptaddress]{revtex4-2}

\usepackage[dvipdfmx]{graphicx}
\usepackage{bm}
\usepackage{amsmath}
\usepackage{amssymb}
\usepackage{newtxtext}
\usepackage{newtxmath}
\usepackage{color}
\usepackage{hyperref}
\usepackage{tikz}
\usepackage{esint}
\usepackage{float} 
\allowdisplaybreaks[1]

\begin{document}

\title{Neural optimization of the most probable paths of 3D active Brownian particles}

\author{Bin Zheng}
\affiliation{Zhejiang Key Laboratory of Soft Matter Biomedical Materials, Wenzhou Institute, 
University of Chinese Academy of Sciences, Wenzhou, Zhejiang 325000, China}

\author{Zhongqiang Xiong}
\affiliation{Zhejiang Key Laboratory of Soft Matter Biomedical Materials, Wenzhou Institute, 
University of Chinese Academy of Sciences, Wenzhou, Zhejiang 325000, China}

\author{Changhao Li}
\affiliation{College of Energy Engineering and State Key Laboratory of Clean Energy Utilization, Zhejiang University, 
Hangzhou, Zhejiang 310003, China}
\affiliation{Joint Research Centre on Medicine, Xiangshan Hospital of Wenzhou Medical University, 
Ningbo, Zhejiang 315700, China}
\affiliation{Zhejiang Key Laboratory of Soft Matter Biomedical Materials, Wenzhou Institute, 
University of Chinese Academy of Sciences, Wenzhou, Zhejiang 325000, China}

\author{Zhanglin Hou}
\affiliation{Zhejiang Key Laboratory of Soft Matter Biomedical Materials, Wenzhou Institute, 
University of Chinese Academy of Sciences, Wenzhou, Zhejiang 325000, China}

\author{Ziluo Zhang}
\affiliation{Department of Physics, Xiamen University, Xiamen, Fujian 361005, China}

\author{Xinpeng Xu}
\affiliation{Department of Physics and MATEC Key Lab, Guangdong Technion -- Israel Institute of Technology, Shantou, Guangdong 515063, China}
\affiliation{Technion - Israel Institute of Technology, Haifa 32000, Israel}

\author{Li-Shing Lin}
\affiliation{Research Institute for Mathematical Sciences, Kyoto University, Kyoto 606-8502, Japan}

\author{Kenta Ishimoto}
\affiliation{Department of Mathematics, Kyoto University, Kyoto 606-8502, Japan}

\author{Kento Yasuda}
\affiliation{Laboratory of Physics, College of Science and Technology, Nihon University, Funabashi, Chiba 274-8501, Japan}

\author{Shigeyuki Komura}\email{Corresponding author: komura@wiucas.ac.cn}
\affiliation{Zhejiang Key Laboratory of Soft Matter Biomedical Materials, Wenzhou Institute, 
University of Chinese Academy of Sciences, Wenzhou, Zhejiang 325000, China}


\begin{abstract}
We develop a variational neural-network framework to determine the most probable path (MPP) of a 3D 
active Brownian particle (ABP) by directly minimizing the Onsager-Machlup integral (OMI).
To obtain the OMI, we use the Onsager-Machlup variational principle for active systems and construct the Rayleighian 
of the ABP by including its active power. 
This approach reveals geometric transitions of the MPP from in-plane I- and U-shaped paths to 3D helical paths as the 
final time and net displacement are varied. 
We also demonstrate that the initial and final boundary conditions have a significant impact on the MPPs.  
Our results show that neural optimization combined with the Onsager-Machlup variational principle provides an efficient 
and versatile framework for exploring optimal transition pathways in active and nonequilibrium systems.
\end{abstract}

\maketitle

\textit{Introduction.--}
Active matter, such as flocks of birds, schools of fish, and bacterial suspensions, has attracted enormous 
interest in nonequilibrium statistical mechanics and biophysics~\cite{LaugaBook,Gompper20}.
A variety of active-particle models have been proposed to investigate emergent collective behaviors.
Active Brownian particles (ABPs) provide a paradigmatic framework to capture the interplay 
between persistent self-propulsion and stochastic fluctuations~\cite{Romanczuk12}.
Typical nonequilibrium phenomena, such as motility-induced phase separation, have been extensively examined 
using interacting ABPs~\cite{Fily12,Cates15}.
Recently, the transition path of an ABP moving in a 2D space between prescribed initial and final positions
was analyzed~\cite{PRE064120}.
Such problems are pertinent to optimal transport in active matter~\cite{Das2024} and have also been 
investigated for active Ornstein-Uhlenbeck particles~\cite{Crisanti23,Dutta23}.

In Ref.~\cite{PRE064120}, by using the Onsager-Machlup variational principle for time-global
processes~\cite{OM1953,Machlup53}, the most probable path (MPP) of an ABP was obtained
through the minimization of the Onsager-Machlup integral (OMI).
Importantly, the OMI is the time integral of the Rayleighian~\cite{PRE063303}, 
which consists of the dissipation function and the time-derivative of the free energy~\cite{OVP284118,DoiBook2013}. 
Various time-local equations for dissipative processes have been systematically derived within the 
Onsager variational principle~\cite{Onsager1931I,Onsager1931II} by using different 
Rayleighians~\cite{Xu15,Okamoto16,Man17,Zhou18,Doi21}. 
Recent works have shown that the Onsager principle can be applied to nonequilibrium active systems by including 
active power in the Rayleighian~\cite{Zhang20,SM3634,EPJP1103}.
Moreover, the Onsager-Machlup principle was also extended to nonequilibrium steady states and further 
connected to fluctuation theorems~\cite{Taniguchi07,Taniguchi08}.

Recent advances in neural-network-based machine learning, particularly physics-informed neural networks, 
provide powerful tools for variational problems that minimize physically meaningful 
quantities~\cite{NRP422,SM5359,Dulaney2021,Colen2021,Wang22,Li25}.
For optimal-transport problems, trajectories characterized by positions and orientations are parameterized 
by neural networks (NNs), and an optimized solution can be obtained by gradient-based training with an appropriate loss function~\cite{Nature1986,Wei2022,PRR033016}. 
This framework allows us to directly minimize the OMI and enables the exploration of MPPs in systems that were 
previously too complex to handle.

In this Letter, we introduce a theoretical framework for efficiently determining the MPP of a 
three-dimensional (3D) ABP under prescribed initial and final boundary conditions.
Our approach has two important aspects:
(i) constructing the OMI from the Rayleighian by incorporating the active power of ABPs~\cite{SM3634},
and (ii) minimizing the OMI via NN-based machine learning~\cite{Li25}.
We show that geometric transitions of the MPP from in-plane I- and U-shaped paths to 3D helical paths appear 
as the final time and the net displacement are varied.
These behaviors arise from competition between orientational constraints and increased configurational 
freedom in 3D. 
The sequence of transitions is significantly influenced by the boundary conditions, although the emergence 
of helical paths is robust.

\begin{figure}[tbh]
{\includegraphics[width=0.4 \textwidth,draft=false]{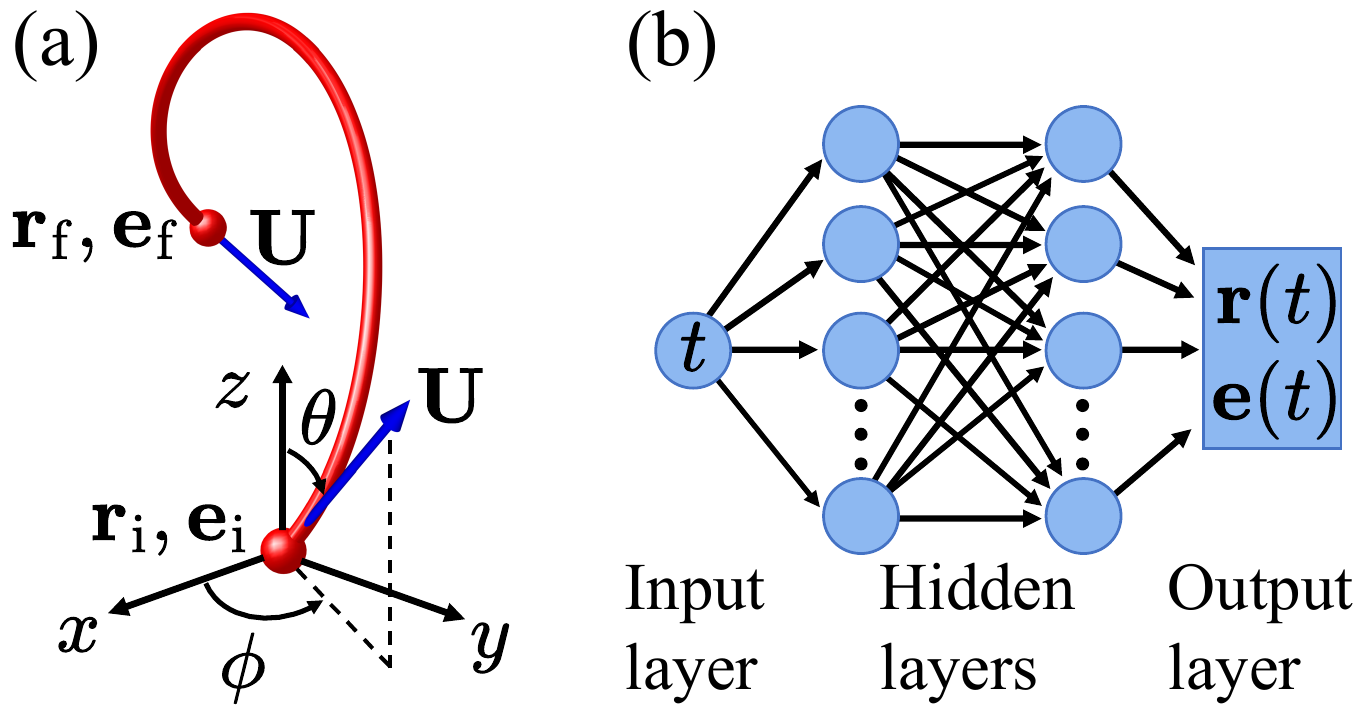}}
\caption{
(a) A trajectory (red curve) of a 3D ABP described by a time-dependent position vector $\mathbf{r}(t)$.
The self-propulsion velocity is $\mathbf{U}(t)=U\mathbf{e}(t)$, where $U$ is a constant  
speed and $\mathbf{e}(t)$ is the orientational unit vector parameterized by the polar angle $\theta$
and azimuthal angle $\phi$. 
The translational and rotational friction coefficients of the ABP are denoted by $\zeta_{\rm t}$ and $\zeta_{\rm r}$, respectively. 
We consider the conditional path probability starting from 
$\mathbf{r}_{\mathrm{i}}$ with $\mathbf{e}_{\mathrm{i}}$ at $t=0$ and 
terminating at $\mathbf{r}_{\mathrm{f}}$ with $\mathbf{e}_{\mathrm{f}}$ at $t=t_{\rm f}$. 
The initial and the final self-propulsion velocities $\mathbf{U}$ are shown by blue arrows. 
(b) Schematic diagram of NN representation of position $\mathbf{r}(t)$ and orientation 
vector $\mathbf{e}(t)$.
The fully connected neural network consists of an input layer, two hidden layers (each with 30 nodes), and output 
layers for the respective functions. 
Activation functions are applied sequentially as tanh (first hidden layer), tanh (second hidden layer), and linear 
(output layer) for $\mathbf{r}(t)$, and tanh (first hidden layer), linear (second hidden layer), 
and linear (output layer) for $\mathbf{e}(t)$.
}
\label{fig1}
\end{figure}

\textit{3D ABP.--}
As shown in Fig.~\ref{fig1}(a), consider an ABP whose translational and rotational friction coefficients are
$\zeta_{\rm t}$ and $\zeta_{\rm r}$, respectively.
The instantaneous 3D position of the particle is $\mathbf{r}(t)=[x(t),y(t),z(t)]$ with translational velocity 
$\mathbf{v}(t)=\dot{\mathbf{r}}(t)=d\mathbf{r}/dt$.
Self-propulsion is introduced by a constant-magnitude velocity $\mathbf{U}(t)=U\mathbf{e}(t)$, where $U$ is a constant 
propulsion speed and $\mathbf{e}(t) = [\sin\theta(t) \cos\phi (t), \sin\theta(t)\sin\phi(t), \cos\theta(t)]$ is the 
orientational unit vector parameterized by the polar and azimuthal angles denoted by $\theta(t)$ and $\phi(t)$, respectively.
The kinematics of $\mathbf{e}$ are given by $\dot{\mathbf{e}}(t)=\bm{\omega}(t)\times\mathbf{e}(t)$, where 
$\bm{\omega}(t)$ is the angular velocity vector~\cite{DoiBook2013}.

The Rayleighian of an ABP is given by  $R = \Phi - \dot{W}_{\mathrm{a}}$, 
where $ \Phi = \zeta_{\rm t}\mathbf{v}^2/2 + \zeta_{\rm r}\bm{\omega}^2/2$ is the dissipation function 
and $\dot{W}_{\mathrm{a}} = \zeta_{\rm t} \mathbf{U} \cdot \mathbf{v}$ is the active power due to the 
self-propulsion~\cite{SM3634}.
In the current problem, no free energy term enters the Rayleighian because we consider a free ABP.
According to the Onsager variational principle~\cite{OVP284118,DoiBook2013}, minimization of $R$ with respect 
to $\mathbf{v}$ and $\bm{\omega}$ yields the most probable kinematics, which are $\mathbf{v}=\mathbf{U}$ and 
$\bm{\omega}=\mathbf{0}$, respectively. 
These are the deterministic equations of motion for a 3D active particle in the absence of thermal fluctuations.
Notice that the minimized Rayleighian (with respect to both $\mathbf{v}$ and $\bm{\omega}$) becomes 
$R_{\min}= - \zeta_{\rm t} U^2/2$.
The Langevin equations of an ABP can be obtained by adding thermal noise to the above deterministic 
equations to describe translational and rotational diffusion~\cite{Romanczuk12}.

\textit{Onsager-Machlup integral.--}
In the presence of Gaussian thermal fluctuations, the ABP executes a stochastic trajectory. 
In an isothermal bath at temperature $T$, the conditional path probability of a trajectory 
starting from $\mathbf{r}_{\mathrm{i}}$ and orientation $\mathbf{e}_{\mathrm{i}}$ at the initial time $t=0$ 
is $P[\mathbf{r}(t),\mathbf{e}(t)|\mathbf{r}_\mathrm{i},\mathbf{e}_\mathrm{i}] = C
\exp \left(-O[\mathbf{r}(t),\mathbf{e}(t)]/(2k_{\rm B}T)\right)$,
where $k_{\rm B}$ is the Boltzmann constant and $C$ is a normalization 
constant~\cite{OM1953,Machlup53}.
The quantity $O$, called the OMI, is constructed as the time integral of the Rayleighian as~\cite{PRE063303}
\begin{align}
O[\mathbf{r}(t),\mathbf{e}(t)] 
& = \int_0^{t_{\rm f}} dt\, (R-R_\mathrm{min})
\label{integralR}
\\
& =\int_0^{t_{\rm f}} dt \, \left[\frac{\zeta_{\rm t}}{2} \left(\mathbf{v}-\mathbf{U}\right)^2 
+ \frac{\zeta_{\rm r}}{2}\bm{\omega}^2\right], 
\label{OMI}
\end{align}
where $t_{\rm f}$ is the final time and Eq.~(\ref{OMI}) is the OMI for the 3D ABP.
Assuming that $\bm{\omega}$ is perpendicular to $\mathbf{e}$, i.e., $\bm{\omega}\cdot\mathbf{e}=0$, we 
have $\bm{\omega}^2 = \dot{\mathbf{e}}^2=\dot{\theta}^2+\dot{\phi}^2 \sin^2\theta$ in terms of spherical 
coordinates~\cite{DoiBook2013}. 
The same OMI can be obtained by using the corresponding Fokker-Planck equation~\cite{PRE064120}.

Defining the characteristic length scale $L=\sqrt{\zeta_{\rm r}/\zeta_{\rm t}}$ and the characteristic time scale 
$L/U$, we use dimensionless quantities such as $\overline{\mathbf{r}} = \mathbf{r}/L$ and $\overline{t}=tU/L$.  
Then the OMI in Eq.~(\ref{OMI}) can be written as 
\begin{align}
\frac{O[\mathbf{r}(t),\mathbf{e}(t)]}{k_{\rm B}T} &= 
\frac{ {\rm Pe}}{2}\int_0^{\overline{t}_{\rm f}}d\overline{t} \, \Big[ 
(\dot{\overline{x}}-\sin\theta\cos\phi)^2 
+ (\dot{\overline{y}}-\sin\theta\sin\phi)^2 
\nonumber \\
& + (\dot{\overline{z}}-\cos\theta)^2 + \dot{\theta}^2+\dot{\phi}^2\sin^2\theta\Big], 
\label{OMI_eq}
\end{align}
where ${\rm Pe} = U/(D_{\rm r} L)$ is the rotational P\'{e}clet number and $D_{\rm r}=k_{\rm B}T/\zeta_{\rm r}$
is the rotational diffusion constant that obeys the Einstein relation~\cite{DoiBook2013}.  
(Hereafter, dot indicates the derivative with respect to the dimensionless time $\overline{t}$.)

Among all stochastic trajectories, the path probability is maximal when the OMI is minimized with respect 
to variations of $\mathbf{r}(t)$ and $\mathbf{e}(t)$.
Hence, the extremal condition $\delta O=0$ determines the MPP, 
$\mathbf{r}^\ast(t)$ and $\mathbf{e}^\ast(t)$, of the 3D ABP.
In the current Langevin bridge problem~\cite{Orland11}, we require that the MPP satisfies the Dirichlet boundary 
conditions at $t=0$ and $t_{\rm f}$, namely,  $\mathbf{r}(0)=\mathbf{r}_{\rm i}$, $\mathbf{e}(0)=\mathbf{e}_\mathrm{i}$ 
and  $\mathbf{r}(t_{\rm f})=\mathbf{r}_{\rm f}$, $\mathbf{e}(t_{\rm f})=\mathbf{e}_\mathrm{f}$.
(A Langevin bridge is a Langevin diffusion conditioned on fixed endpoints, which is related to 
the spontaneous folding of proteins and allosteric transitions~\cite{Orland11, LeTreut25}.) 
The nonlinear Euler-Lagrange equations obtained from Eq.~(\ref{OMI_eq}) are 
\begin{eqnarray}
& \frac{d}{d\overline{t}}(\dot{\overline{x}}-\sin\theta\cos\phi) =  
\frac{d}{d\overline{t}}(\dot{\overline{y}}-\sin\theta\sin\phi) = 
\frac{d}{d\overline{t}}(\dot{\overline{z}}-\cos\theta) = 0, \nonumber \\
& \dot{\overline{x}}\cos\theta\cos\phi+\dot{\overline{y}}\cos\theta\sin\phi -\dot{\overline{z}}\sin\theta-\dot{\phi}^2\sin\theta\cos\theta+\ddot{\theta} = 0, \nonumber \\
& \dot{\overline{x}}\sin\phi- \dot{\overline{y}}\cos\phi-2\dot{\theta}\dot{\phi}\cos\theta-\ddot{\phi}\sin\theta = 0. 
\label{ELequation}
\end{eqnarray}
In contrast to the 2D case~\cite{PRE064120}, these equations for a 3D ABP are difficult to solve analytically because they are 
more general than those of a 3D spherical pendulum that satisfies the angular momentum conservation. 
In general, the Euler-Lagrange equations provide only a necessary condition for an extremum, and there is no guarantee 
that the extremum is a minimum.

According to local detailed balance~\cite{Sekimoto11,Peliti2021,Shiraishi2023,Seifert2025}, the entropy change 
$\Delta S[\mathbf{r}(t),\mathbf{e}(t)]$ satisfies the relation
\begin{equation}
e^{\Delta S[\mathbf{r}(t),\mathbf{e}(t)]/k_{\rm B} }   
= \frac{P[\mathbf{r}(t),\mathbf{e}(t)|\mathbf{r}_\mathrm{i},\mathbf{e}_\mathrm{i}]}
{P[\mathbf{r}^\text{rev}(t),\mathbf{e}^\text{rev}(t)|\mathbf{r}_\mathrm{f},\mathbf{e}_\mathrm{f}]}, 
\label{FluctuationTheorem}
\end{equation}
where $\mathbf{r}^\text{rev}(t) = \mathbf{r}(t_{\rm f}-t)$ and $\mathbf{e}^\text{rev}(t) = \mathbf{e}(t_{\rm f}-t)$
represent the reversed paths. 
Physically, the quantity $T \Delta S$ is the heat dissipated to the thermal bath.
For the 3D ABP, the entropy change can be written as  
\begin{align}
\frac{\Delta S[\mathbf{r}(t),\mathbf{e}(t)]}{k_{\rm B}} = {\rm Pe} \int_0^{\overline{t}_{\rm f}}d\overline{t} \, \left[ \dot{\overline{x}}\sin\theta\cos\phi + \dot{\overline{y}}\sin\theta\sin\phi \right.
\nonumber \\
\left. + \dot{\overline{z}}\cos\theta \right],
\label{entropy}
\end{align}
which will later be evaluated along the MPP, $\mathbf{r}^\ast(t)$ and $\mathbf{e}^\ast(t)$.
Notice that both Eqs.~(\ref{OMI_eq}) and (\ref{entropy}) are proportional to the P\'{e}clet number, 
which is proportional to the self-propulsion speed $U$, and hence they vanish in the passive limit.
The corresponding expressions for 2D ABPs can be obtained by setting $\theta=\pi/2$~\cite{PRE064120}.

\begin{figure*}[htb]
{\includegraphics[width=0.9\textwidth,draft=false]{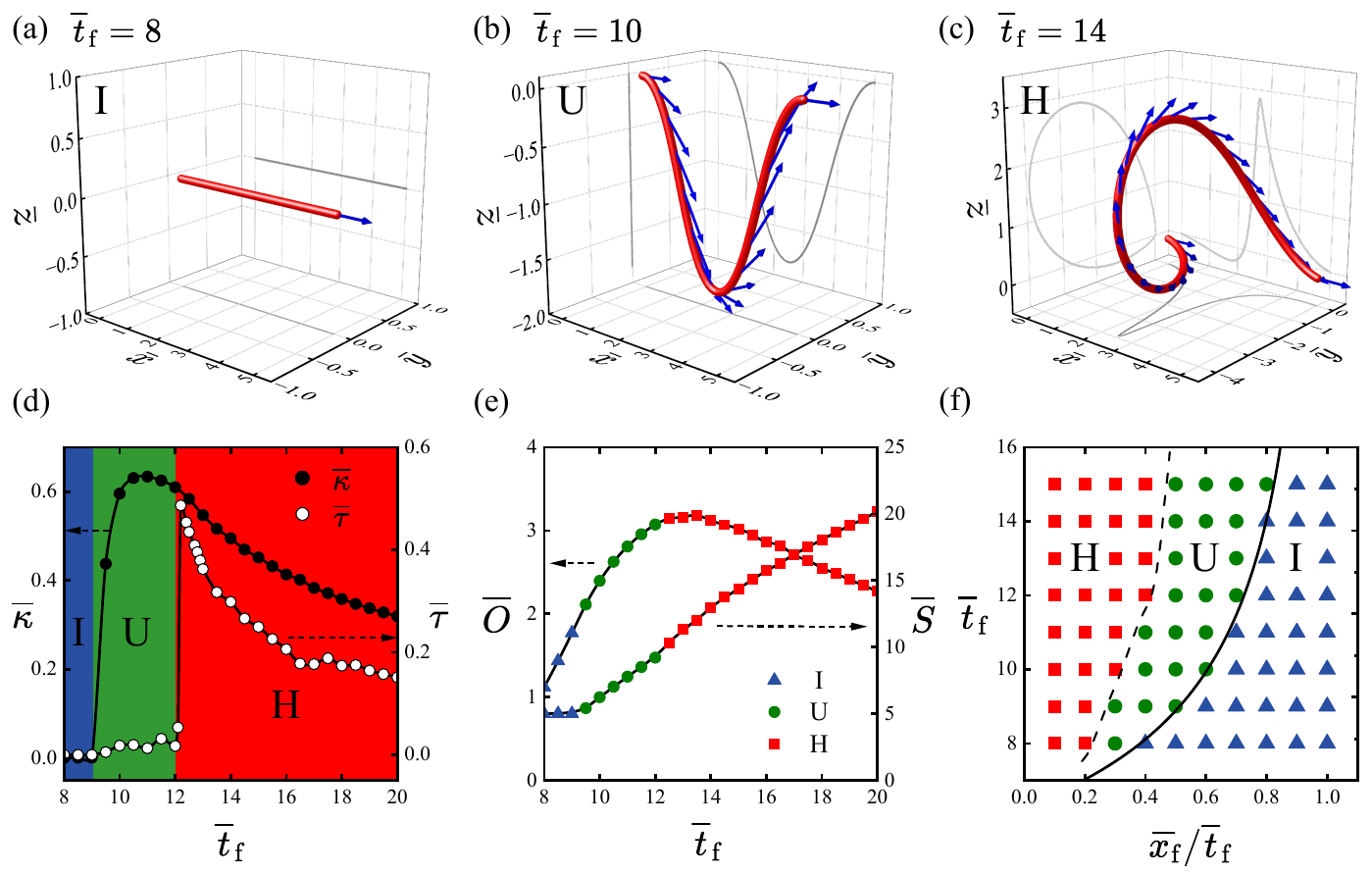}}
\caption{
NN solutions of the MPP of a 3D ABP for different final times $\overline{t}_{\rm f}$, 
with fixed initial and final boundary conditions
$\overline{x}_{\rm i}=\overline{y}_{\rm i}=\overline{z}_{\rm i}=\phi_{\rm i}=0$, 
$\theta_{\rm i}=\pi/2$ and 
$\overline{x}_{\rm f}=5$, $\overline{y}_{\rm f}=\overline{z}_{\rm f}=\phi_{\rm f}=0$, 
$\theta_{\rm f}=\pi/2$, respectively.
(a) I-path when $\overline{t}_{\rm f}=8$,  
(b) U-path when $\overline{t}_{\rm f}=10$,
and (c) H-path when $\overline{t}_{\rm f}=14$.
The red curves represent the MPPs $\mathbf{r}^\ast(t)$,  
and the gray curves show their projections onto the $x$-$y$, $x$-$z$, and $y$-$z$ planes. 
The blue arrows indicate the optimized orientations $\mathbf{e}^\ast(t)$ of the ABP. 
(d) The average curvature $\overline{\kappa}$ (left axis) and the torsion $\overline{\tau}$ (right axis) as  
functions of the final time $\overline{t}_{\rm f}$ (see Eq.~(\ref{geometry}) and the text).
The blue, green, and red background regions correspond to I-, U-, and H-paths, respectively.
(e) The dimensionless OMI $\overline{O} = 2 O/(k_{\rm B}T {\rm Pe})$ [left axis, see Eq.~(\ref{OMI_eq})] and 
the entropy change $\overline{S}=\Delta S/(k_{\rm B}{\rm Pe})$ [right axis, see Eq.~(\ref{entropy})] as 
functions of the final time $\overline{t}_{\rm f}$.
The blue triangles, green circles, and red squares represent I-, U-, and H-paths, respectively.
(f) Phase diagram of the MPPs in the parameter space spanned by 
$\overline{x}_{\rm f}/\overline{t}_{\rm f}$ and $\overline{t}_{\rm f}$.
The symbols are the same as in (e).
The solid black line represents the analytical line (see the text) that separates the I- and U-paths in 2D, 
while the black dashed line is a guide for the eye.  
} 
\label{fig2}
\end{figure*}

\begin{figure*}[htb]
{\includegraphics[width=0.9\textwidth,draft=false]{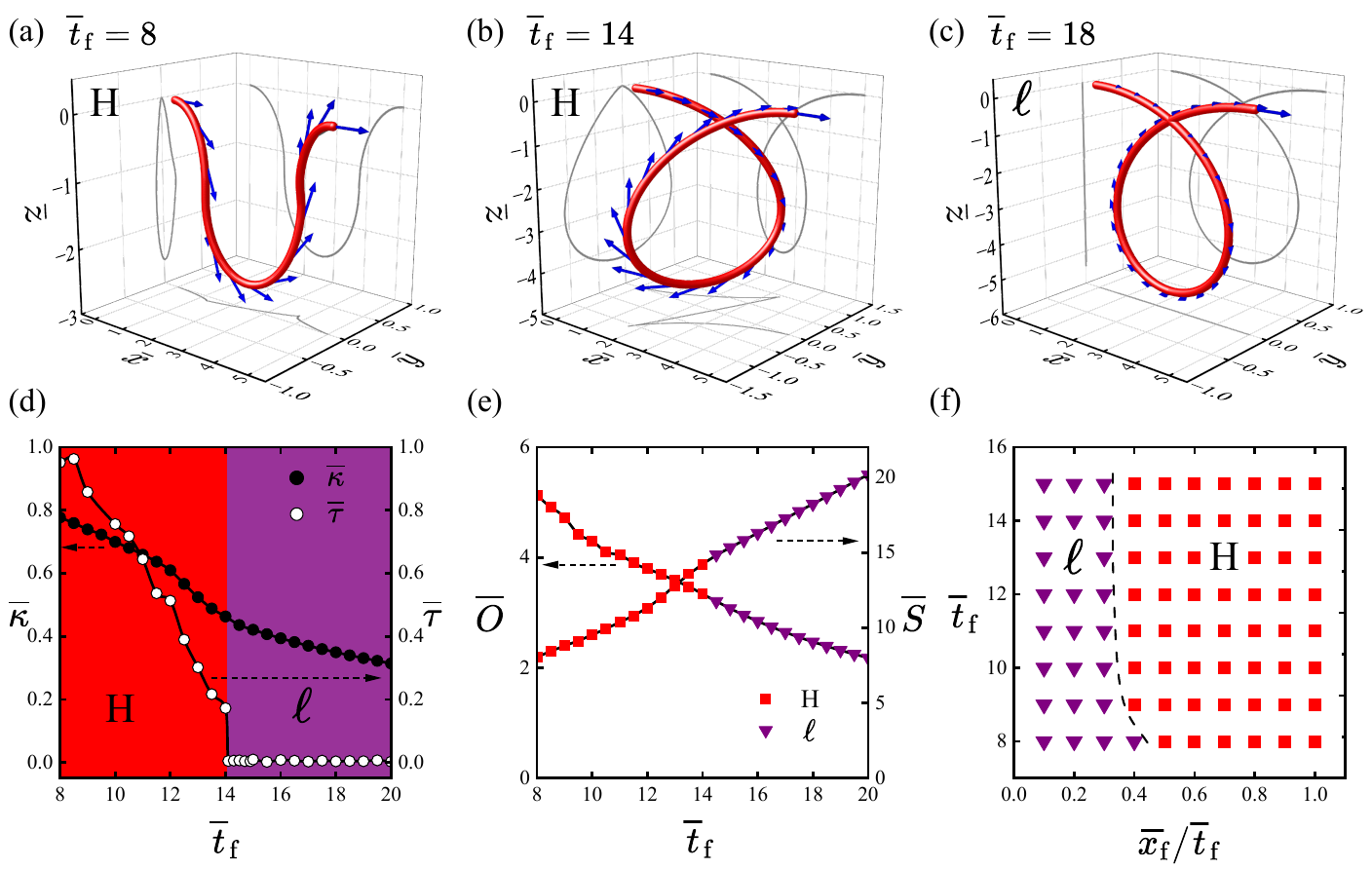}}
\caption{
NN solutions of the MPP of a 3D ABP for different final times $\overline{t}_{\rm f}$, 
with fixed initial and final boundary conditions
$\overline{x}_{\rm i}=\overline{y}_{\rm i}=\overline{z}_{\rm i}=\phi_{\rm i}=0$, $\theta_{\rm i}=\pi/2$ 
and $\overline{x}_{\rm f}=5$, $\overline{y}_{\rm f}=\overline{z}_{\rm f}=\phi_{\rm f}=0$, 
$\theta_{\rm f}=5\pi/2$, respectively.
(a) H-path when $\overline{t}_{\rm f}=8$,  
(b) H-path when $\overline{t}_{\rm f}=14$,
and (c) $\ell$-path when $\overline{t}_{\rm f}=18$.
(d) The average curvature $\overline{\kappa}$ (left axis) and the torsion $\overline{\tau}$ (right axis) as  
functions of the final time $\overline{t}_{\rm f}$ (see Eq.~(\ref{geometry}) and the text).
The red and purple background regions correspond to H- and $\ell$-paths, respectively.
(e) The dimensionless OMI $\overline{O} = 2 O/(k_{\rm B}T {\rm Pe})$ [left axis, see Eq.~(\ref{OMI_eq})] and 
the entropy change $\overline{S}=\Delta S/(k_{\rm B}{\rm Pe})$ [right axis, see Eq.~(\ref{entropy})] 
as functions of the final time $\overline{t}_{\rm f}$.
The red squares and purple triangles represent H- and $\ell$-paths, respectively.
(f) Phase diagram of the MPPs in the parameter space spanned by 
$\overline{x}_{\rm f}/\overline{t}_{\rm f}$ and $\overline{t}_{\rm f}$.
The black dashed line is a guide for the eye. 
}
\label{fig3}
\end{figure*}

\textit{Minimization of the OMI by NN.--}
Rather than solving the Euler-Lagrange equations, we minimize the OMI directly using machine learning to obtain the MPP. 
The position $\mathbf{r}(t)$ and orientation $\mathbf{e}(t)$ are represented by fully connected NNs, as schematized in 
Fig.~\ref{fig1}(b). 
Each NN takes time $\overline{t}$, ranging over $[0,\overline{t}_{\rm f}]$ with a time step of $10^{-3}$, as input and 
passes through two hidden layers, followed by a linear output layer.
To provide sufficient nonlinearity without overconstraining the outputs, the position networks use 
hyperbolic tangent ($\tanh$) activations in both hidden layers and a linear output, whereas the orientation networks apply 
$\tanh$  only in the first hidden layer, followed by a linear second hidden layer and a linear output layer.

The minimization problem is converted to the optimization of these five NNs under the loss function
\begin{equation}
J =\overline{O}+\sum_{\mathbf{K}=\mathbf{r},\mathbf{e}}w_{\rm i}\left(\mathbf{K}_{\rm i}^{\rm NN}-\mathbf{K}_{\rm i}\right)^2
+\sum_{\mathbf{K}=\mathbf{r},\mathbf{e}}w_{\rm f}\left(\mathbf{K}_{\rm f}^{\rm NN}-\mathbf{K}_{\rm f}\right)^2,
\end{equation}
where $\overline{O} = 2 O/(k_{\rm B}T {\rm Pe})$ and 
$\mathbf{K}$ denotes the position $\overline{\mathbf{r}}$ and orientation $\mathbf{e}$ with 
$\mathbf{K}^{\mathrm{NN}}$ representing the values given by the NN, and unsuperscripted $\mathbf{K}$ denoting the 
target boundary condition values.
The OMI in Eq.~(\ref{OMI_eq}) has a quadratic form, and the remaining terms enforce boundary 
conditions with weights $w_{\rm i}=w_{\rm f}=100$. 
The time-derivative contributions in $O$ are computed via the automatic differentiation function (torch.autograd.grad)
in PyTorch, and the Adam optimization algorithm is implemented to update the network parameters. 
After training, the optimized networks yield the MPP, $\mathbf{r}^\ast(t)$ and $\mathbf{e}^\ast(t)$.

During network training, we tested different network architectures, including three and four hidden layers, with the number of 
neurons in each layer varied from 30 to 60. 
We also examined the time discretization in the range 10$^{-4}$ to 10$^{-3}$ and tested the penalty weights for the boundary 
conditions with values 1, 100, and 500.
Based on these tests, we found that, within a suitable parameter range, the obtained MPPs and OMI values show no significant quantitative changes.
We have also verified numerically that the obtained MPPs satisfy the Euler-Lagrange equations in Eq.~(\ref{ELequation}).

We present the NN results for the MPP under the following conditions. 
The ABP starts from the initial state with $\overline{x}_{\rm i}=\overline{y}_{\rm i}=\overline{z}_{\rm i}=\phi_{\rm i}=0$, 
$\theta_{\rm i}=\pi/2$ and ends at the final position 
$\overline{x}_{\rm f}=5$, $\overline{y}_{\rm f}=\overline{z}_{\rm f}=\phi_{\rm f}=0$, $\theta_{\rm f}=\pi/2$ 
or $5\pi/2$ chosen to examine the effect of different final orientation boundary conditions.
In contrast to the 2D ABP case~\cite{PRE064120}, the additional spatial degree of freedom in 3D gives rise to a new 
trajectory type, as explained below.

\textit{Case $\theta_{\rm f}=\pi/2$.--}
In Figs.~\ref{fig2}(a)-(c), we show three representative MPPs (red trajectories) obtained for different 
final times $\overline{t}_{\mathrm{f}}$. 
For $\overline{t}_{\mathrm{f}}=8$, the MPP is a straight in-plane trajectory, called 
the I-path, whose projection onto the $y$-$z$ plane collapses to a single point [Fig.~\ref{fig2}(a)]. 
As the final time is increased to $\overline{t}_{\mathrm{f}}=10$, the MPP becomes curved in the $x$-$z$ plane 
[Fig.~\ref{fig2}(b)], called the U-path. 
For a larger final time $\overline{t}_{\mathrm{f}}=14$, the MPP evolves into a 3D helical trajectory, called the 
H-path, as evidenced by its circular projection in the $y$-$z$ plane [Fig.~\ref{fig2}(c)].

To quantitatively characterize transitions among different types of paths, we compute the time-dependent 
curvature $\kappa(t)$ and torsion $\tau(t)$ of the MPP via the standard differential-geometric 
definitions~\cite{AudolyBook}:
\begin{align}
\kappa(t) = \frac{\left\vert \dot{\mathbf{r}} \times \ddot{\mathbf{r}} \right\vert}
{\left\vert \dot{\mathbf{r}} \right\vert^{3}},
\quad
\tau(t) = \frac{\big(\dot{\mathbf{r}}\times \ddot{\mathbf{r}}\big)\cdot \dddot{\mathbf{r}}}
{\left\vert \dot{\mathbf{r}}\times \ddot{\mathbf{r}} \right\vert^{2}}.
\label{geometry}
\end{align}
For convenience, we further define trajectory-averaged curvature and torsion as 
$\overline{\kappa} = (L/t_{\rm f}) \int_0^{t_{\rm f}} dt\, \kappa(t)$ 
and 
$\overline{\tau} = (L/t_{\rm f}) \int_0^{t_{\rm f}} dt\, \vert \tau(t) \vert$, 
respectively, to characterize the overall structure of the MPPs. 
In the latter quantity, we integrate the absolute value of the torsion $\vert \tau(t) \vert$.

In Fig.~\ref{fig2}(d), we show both $\overline{\kappa}$ (filled circles) and $\overline{\tau}$ (open circles)
as functions of the final time $\overline{t}_{\rm f}$.
For small $\overline{t}_{\mathrm f}$, both $\overline{\kappa}$ and $\overline{\tau}$ vanish within numerical 
precision, corresponding to the I-path (blue background region).
For $\overline{t}_{\mathrm f} \gtrsim 9$, $\overline{\kappa}$ increases while $\overline{\tau}$ 
remains nearly zero, consistent with the U-path (green background region).
For $\overline{t}_{\mathrm f} \gtrsim 12$, the torsion increases significantly, indicating the emergence 
of the H-path (red background region).
For the H-path, both the curvature and torsion decrease monotonically as $\overline{t}_{\mathrm f}$ increases.

For the obtained MPPs, we plot in Fig.~\ref{fig2}(e) the dimensionless OMI, 
$\overline{O} = 2 O/(k_{\rm B}T {\rm Pe})$ [see Eq.~(\ref{OMI_eq})], 
and the entropy change, 
$\overline{S}=\Delta S/(k_{\rm B}{\rm Pe})$ [see Eq.~(\ref{entropy})], as functions 
of the final time $\overline{t}_{\mathrm f}$.
Different types of paths identified in Fig.~\ref{fig2}(d) are represented by different symbols.
The minimized OMI values for the I- and U-paths coincide with the analytical results reported for a 2D ABP 
in Ref.~\cite{PRE064120}, thereby validating our NN-based computations.
For the H-paths ($\overline{t}_{\mathrm f}>12$), we find that the OMI generally decreases as $\overline{t}_{\mathrm f}$ 
is increased. 
This is because the additional out-of-plane degrees of freedom in 3D permit trajectories with a lower OMI 
than in 2D.
Regarding the entropy change, $\overline{S}$ remains constant for the I-path and is given by 
$\overline{x}_{\rm f}$, which is equal to $5$ in this case. 
Beyond this regime, $\overline{S}$ monotonically increases with $\overline{t}_{\rm f}$, indicating that the time-reversed 
process is less likely to follow the same path for larger $\overline{t}_{\mathrm f}$.

Figure~\ref{fig2}(f) depicts the phase diagram of the MPPs in the parameter space spanned by 
$\overline{x}_{\rm f}/\overline{t}_{\rm f}$ and $\overline{t}_{\rm f}$.
Note that the former quantity corresponds to the dimensionless apparent velocity of the ABP.
For a fixed $\overline{t}_{\rm f}$, the MPP transforms from the H-path to the U-path and further 
to the I-path as $\overline{x}_{\rm f}/\overline{t}_{\rm f}$ is increased. 
The U-path region becomes narrower as $\overline{t}_{\rm f}$ decreases. 
The analytical (black solid) line, $\overline{t}_{\rm f} = 2\pi /\sqrt{1-\overline{x}_{\rm f}/\overline{t}_{\rm f}}$, 
separating the regions of the U-path and the I-path was derived for a 2D ABP~\cite{PRE064120}, and agrees 
well with our NN results.

\textit{Case $\theta_{\rm f}=5\pi/2$.--}
Next, we examine the effect of the periodicity of $\theta$ by setting $\theta_{\rm f}=\pi/2+2\pi=5\pi/2$, 
which corresponds to an additional $2\pi$ rotation of $\theta$.
The other boundary conditions remain the same as before. 
In Figs.~\ref{fig3}(a)-(c), we show three representative MPPs obtained for different final times 
$\overline{t}_{\mathrm{f}}$. 
For $\overline{t}_{\rm f}=8$, the MPP forms a twisted U-shaped path in 3D space [Fig.~\ref{fig3}(a)], which is 
classified as an H-path because a small annulus appears in its projection onto the $y$-$z$ plane.
As the final time is increased to $\overline{t}_{\rm f}=14$, the optimal trajectory becomes a fully developed H-path
[Fig.~\ref{fig3}(b)].
When $\overline{t}_{\mathrm{f}}=18$, however, the optimized trajectory forms an in-plane $\ell$-path because the
ABP must undergo a $2\pi$ rotation in $\theta$ [Fig.~\ref{fig3}(c)].

In Fig.~\ref{fig3}(d), we show both the average curvature $\overline{\kappa}$ and torsion $\overline{\tau}$ as functions
of $\overline{t}_{\mathrm{f}}$ when $\theta_{\rm f}=5\pi/2$.
In contrast to Fig.~\ref{fig2}(d) for $\theta_{\rm f}=\pi/2$, both $\overline{\kappa}$ and $\overline{\tau}$ are 
nonzero for small $\overline{t}_{\mathrm{f}}$. 
When $\overline{t}_{\rm f}=14$, the torsion decreases to zero while the curvature is still nonzero, indicating 
the emergence of the $\ell$-path.
In Fig.~\ref{fig3}(e), we plot the dimensionless OMI $\overline{O}$ and the entropy change $\overline{S}$ as functions 
of $\overline{t}_{\mathrm{f}}$.
Here, the H-path and $\ell$-path are shown by red squares and purple triangles, respectively. 
In contrast to the case shown in Fig.~\ref{fig2}(e), the OMI decreases monotonically as $\overline{t}_{\rm f}$ increases, 
while the entropy change increases as before. 
Although we do not directly compare the OMI values for $\theta_{\rm f}=\pi/2$ and $5\pi/2$, 
they are close to each other when $\overline{t}_{\mathrm{f}}$ is large. 
In Fig.~\ref{fig3}(f), we show the phase diagram of the MPPs in terms of $\overline{x}_{\rm f}/\overline{t}_{\rm f}$ 
and $\overline{t}_{\rm f}$ when $\theta_{\rm f}=5\pi/2$.
In this case, the transition line separating the $\ell$-path and the H-path is almost independent of $\overline{t}_{\rm f}$ 
and is approximately given by $\overline{x}_{\rm f}/\overline{t}_{\rm f} \approx 0.4$.
These results indicate that the boundary conditions have a significant influence on the geometry of the 
MPPs in 3D.

\textit{Summary and discussion.--}
In summary, we have developed a variational neural-network framework to determine the MPP of a 3D ABP 
by directly minimizing the OMI using an NN parametrization. 
Here, the OMI is constructed from the Rayleighian by incorporating the active power of an ABP.
Our results reveal geometric transitions of the MPP,  from in-plane I- and U-shaped paths to 3D helical
H-paths, as the final time and net displacement are varied. 
We have also demonstrated that boundary conditions play a crucial role in selecting the MPP geometry. 
Our study establishes a systematic approach to uncovering optimal transition pathways in active systems in 
higher dimensions and can be extended to study other fluctuation-driven rare events in more complex systems.

One of the central aims of this work is to demonstrate that the Onsager  
principle~\cite{Onsager1931I,Onsager1931II} and the Onsager-Machlup principle~\cite{OM1953,Machlup53} 
can be extended to nonequilibrium active systems. 
Our key idea is to incorporate the active power $\dot{W}_{\mathrm{a}}$ supplied by active forces into the 
Rayleighian~\cite{OVP284118,DoiBook2013}. 
This is possible despite the fact that the active force is not derived from a free energy.
In active systems, the free energy in the Rayleighian should be understood as a non-equilibrium free energy~\cite{Shiraishi2023} 
that represents the passive thermodynamic part of the coarse-grained state, while the genuinely active driving 
is incorporated separately through the active power~\cite{Yasuda26}.
For an ABP, the active power is given by $\dot{W}_{\mathrm{a}} = \zeta_{\rm t} \mathbf{U} \cdot \mathbf{v}$, 
and other examples of active forces can be found in Refs.~\cite{Zhang20,SM3634,EPJP1103}. 
We also note that nonreciprocal interactions~\cite{Fruchart21,Fruchart22,You2020,Saha2020,Liu2023,Tateyama2024} 
can be formulated through the notion of active power~\cite{Yasuda2022,Yasuda2022JCP,Yasuda2024} 
and thus incorporated into the Rayleighian. 
Once the active Rayleighian is constructed, the OMI follows directly from Eq.~(\ref{integralR}).

By directly minimizing the OMI, we have determined the MPP under the given boundary conditions. 
To further discuss the effects of thermal fluctuations, one must employ the statistical formulation of 
the Onsager-Machlup principle, as detailed in Ref.~\cite{PRE044104}.
In this formulation, we introduce a stochastic observable and consider a modified (shifted) OMI that is 
maximized to obtain the cumulant generating function of the observable. 
Such a statistical treatment successfully describes the fluctuating dynamics of fluids under both equilibrium
and nonequilibrium conditions~\cite{PRE044104} and has been further applied to informational active 
matter~\cite{Yasuda2025}. 
Our future task is to maximize the modified OMI of an ABP by using the NN approach to investigate the 
effects of thermal fluctuations.

While we have determined the phase diagrams of the MPPs in Figs.~\ref{fig2}(f) and \ref{fig3}(f), the 
nature of the corresponding transitions merits further investigation. 
For example, the transitions I$\rightarrow$U and U$\rightarrow$H are both discontinuous with respect 
to the curvature and torsion [see Fig.~\ref{fig2}(d)]. 
The transition from in-plane to helical MPPs appears analogous to symmetry breaking in elastic rods.
The connection between 3D optimal paths in active systems and Euler buckling or the elastica theory of 
elastic rods under compression~\cite{AudolyBook} will be addressed in future work.

Machine learning (ML) techniques have recently been applied to a range of active-matter problems, such as 
motility-induced phase separation in ABPs and active-nematic hydrodynamics~\cite{Dulaney2021,Colen2021}.
The variational neural-network framework circumvents the difficulties associated with building 
analytical theories and the numerical challenges encountered in directly solving high-dimensional equations.
In conventional methods, one must search for multiple solution branches and then identify the MPP by 
comparing the corresponding OMI values. 
The current neural-network framework avoids this complicated procedure and directly yields the MPP.
Our approach enables the systematic exploration of MPPs in active systems and can be extended to 
rare-event problems such as optimal transport of ABPs between distinct potentials in 3D space~\cite{Das2024}.
Combining ML optimization with the Onsager-Machlup principle offers a general strategy for studying 
nonequilibrium processes in active matter, providing a direct bridge between microscopic dynamics and 
emergent large-scale behaviors.

\textit{Acknowledgments.--}
K.Y.\ thanks Uwe Thiele and Fr\'ed\'eric van Wijland for useful discussions.
B.Z.\ acknowledges the National Natural Science Foundation of China (Grant No.\ 22203022) and 
the Scientific Research Starting Foundation of Wenzhou Institute, UCAS (Grant No.\ WIUCASQD2022016). 
K.I.\ acknowledges  JSPS KAKENHI for Transformative Research Areas (Grant No.\ 21H05309), 
JST, FOREST, Japan (Grant No.\ JPMJFR212N), and  JST, CREST, Japan (Grant No.\ JPMJCR25Q1).
K.Y.\ acknowledges JSPS KAKENHI for Grant-in-Aid for Early-Career Scientists (Grant No.\ 25K17357).
S.K.\ acknowledges the support from the National Natural Science Foundation of China (Grant No.\ 12274098) and 
from the Zhejiang Key Laboratory of Soft Matter Biomedical Materials (2025ZY01036 and 2025E10072).
K.I., K.Y., and S.K.\ acknowledge the support from the Japan Society for the Promotion of Science (JSPS) Core-to-Core 
Program ``Advanced Core-to-Core network for the physics of self-organizing active matter" (No.\ JPJSCCA20230002).
B.Z.\ and Z.X.\ contributed equally to this work.


\end{document}